\documentclass[11pt]{article}

\usepackage[T1]{fontenc}
\usepackage[utf8]{inputenc}
\usepackage{mathptmx}
\usepackage{microtype}
\usepackage[a4paper,margin=1in]{geometry}
\usepackage{amsmath,amssymb,amsthm,mathtools}
\usepackage{booktabs,multirow,tabularx,array,makecell}
\usepackage{graphicx}
\usepackage{xcolor}
\usepackage{tikz}
\usetikzlibrary{positioning,arrows.meta,calc,shapes.misc}
\usepackage[ruled,vlined,linesnumbered]{algorithm2e}
\usepackage{enumitem}
\usepackage[numbers,sort&compress]{natbib}
\usepackage[hidelinks]{hyperref}
\usepackage[capitalise,noabbrev]{cleveref}
\usepackage{caption}
\setlist{itemsep=2pt,topsep=4pt}

\newtheorem{definition}{Definition}
\newtheorem{proposition}{Proposition}
\newtheorem{corollary}{Corollary}
\theoremstyle{remark}
\newtheorem{remark}{Remark}

\definecolor{fixedc}{HTML}{CFE3F2}
\definecolor{presentc}{HTML}{F6D3C2}
\definecolor{undetc}{HTML}{E4E4E4}
\definecolor{accent}{HTML}{C2571A}
\newcommand{\Pres}{\textup{\textsc{Present}}}
\newcommand{\Fix}{\textup{\textsc{Fixed}}}
\newcommand{\Und}{\textup{\textsc{Undetermined}}}
\newcommand{\reval}{\textup{\textsc{Reval}}}
\newcommand{\FRR}{\mathrm{FRR}}

\newcommand{\nConcerns}{441}
\newcommand{\nFamilies}{12}
\newcommand{\nAssess}{1284}
\newcommand{\nPresent}{631}
\newcommand{\nFixed}{561}
\newcommand{\nUndet}{92}
\newcommand{\nRecur}{155}
\newcommand{\nRecurH}{128}
\newcommand{\nScen}{13}
\newcommand{\nPresentH}{639}
\newcommand{\nFixedH}{553}
\newcommand{\revPerConcern}{2.91}
\newcommand{\nBoot}{2{,}000}
\newcommand{\nDraws}{5}

\newcommand{\truthAgree}{92.8}
\newcommand{\nTruthDisagree}{92}
\newcommand{\FrrOut}{0.73}
\newcommand{\FrrOutLo}{0.66}
\newcommand{\FrrOutHi}{0.78}
\newcommand{\FrrOutPct}{72.9}

\newcommand{\PrecOut}{0.53}

\newcommand{\FrrPat}{0.37}

\newcommand{\FrrOnceK}{0.30}

\newcommand{\CostAlwaysK}{8.73}

\newcommand{\FrrRevalS}{0.37}

\newcommand{\FrrRevalK}{0.11}
\newcommand{\FrrRevalKLo}{0.06}
\newcommand{\FrrRevalKHi}{0.16}

\newcommand{\FrrRevalKDrawLo}{0.10}
\newcommand{\FrrRevalKDrawHi}{0.12}
\newcommand{\RecRevalK}{0.80}
\newcommand{\RecRevalKLo}{0.73}
\newcommand{\RecRevalKHi}{0.88}
\newcommand{\RecRevalKPct}{80.4}

\newcommand{\CostRevalK}{5.73}

\newcommand{\FrrRevalCaH}{0.36}

\newcommand{\FrrOnceCcH}{0.32}

\newcommand{\FrrRevalCcH}{0.13}
\newcommand{\FrrRevalCcHLo}{0.00}
\newcommand{\FrrRevalCcHHi}{0.26}

\newcommand{\RecRevalCcH}{0.88}

\newcommand{\RecRevalCcHPct}{88.3}

\newcommand{\MissRevalCcH}{0.00}

\newcommand{\FrrRevalKH}{0.17}

\newcommand{\FrrRevalCcO}{0.11}

\newcommand{\DFrrOnce}{-0.20}
\newcommand{\DFrrOnceLo}{-0.22}
\newcommand{\DFrrOnceHi}{-0.18}

\newcommand{\DRecFunc}{0.07}
\newcommand{\DRecFuncLo}{0.00}
\newcommand{\DRecFuncHi}{0.14}
\newcommand{\DCostFunc}{0.06}

\newcommand{\DFrrOnceC}{-0.19}
\newcommand{\DFrrOnceCLo}{-0.24}
\newcommand{\DFrrOnceCHi}{-0.15}

\newcommand{\DFrrSingleC}{-0.23}
\newcommand{\DFrrSingleCLo}{-0.31}
\newcommand{\DFrrSingleCHi}{-0.14}

\newcommand{\agreeRevalAlways}{100.0}
\newcommand{\agreeRevalAlwaysCmt}{100.0}
\newcommand{\agreeFuncAlways}{99.2}
\newcommand{\nCompared}{6{,}420}
\newcommand{\reassUnrelReval}{5.0}
\newcommand{\reassUnrelFile}{58.5}
\newcommand{\outClosedPresent}{460}
\newcommand{\outOpenFixed}{41}
\newcommand{\outDivergeConcerns}{66.7}
\newcommand{\costSaving}{34}
\newcommand{\sweepFrra}{0.366}
\newcommand{\sweepReca}{0.45}

\newcommand{\sweepFrrb}{0.205}

\newcommand{\sweepFrrc}{0.106}

\newcommand{\sweepFrrd}{0.058}

\newcommand{\sweepFrre}{0.026}
\newcommand{\sweepRece}{0.97}

\newcommand{\sweepSaveMin}{34}
\newcommand{\sweepSaveMax}{34}

\newcommand{\CatPartPat}{0.93}

\newcommand{\CatHelpPat}{0.92}

\newcommand{\CatConstFuncK}{1.00}

\newcommand{\CatConstRevalK}{0.17}
\newcommand{\nCatCos}{194}
\newcommand{\nCatPart}{145}
\newcommand{\nCatInl}{49}
\newcommand{\nCatHelp}{93}
\newcommand{\nCatConst}{13}
\newcommand{\nCatUnt}{137}

\newcommand{\perExecUntraced}{44}
\newcommand{\perExecTraced}{122}
\newcommand{\traceFactor}{2.8}
\newcommand{\breakEven}{0.42}
\newcommand{\revalOverhead}{1.27}
\newcommand{\timeRatio}{3.3}
\newcommand{\timingRepeats}{7}
\newcommand{\pyVersion}{3.13.16}

\title{\textbf{AI Code Review Requires Revision-Aware\\ Evidence of Concern Resolution}}
\author{Mamdouh Alenezi\\[2pt]
\small SDAIA Academy, Saudi Data and Artificial Intelligence Authority (SDAIA)\\
\small Riyadh, Saudi Arabia}
\date{October 2026}

\begin{document}
\maketitle

% ================================================================== ABSTRACT
\begin{abstract}
\noindent AI-assisted code review increasingly involves automated judgements about whether reported concerns have been resolved. However, administrative closure and changes to commented lines do not establish substantive repair, and subsequent revisions may invalidate an initially correct resolution. This position paper argues that resolution judgements require inspectable evidence tied to a specific code revision and revalidation when changes affect the dependencies supporting that evidence. We formalise review concerns as persistent verification obligations, establish a sufficient condition for dependency-gated revalidation to preserve the decisions of exhaustive reassessment, and instantiate the approach as \reval{}, a lightweight evidence-maintenance procedure. The contribution adapts established principles from regression-test selection and build systems to the longitudinal assessment of individual review concerns. A controlled study evaluates synthetic revision histories comprising \nConcerns{} concerns, \nAssess{} assessments, and \nFamilies{} defect families against oracle ground truth. An outdated-line proxy that retains resolution status falsely classified \FrrOutPct\% of states in which concerns remained present as resolved, while a policy that stopped checking after first success detected no recurrences. With three oracle-sampled witnesses, \reval{} achieved a false-resolution rate of \FrrRevalK{} (95\% CI: \FrrRevalKLo--\FrrRevalKHi) and detected \RecRevalKPct\% of recurrences. With witnesses constructed from review comments alone and evaluated against independently written held-out checks, the corresponding results were \FrrRevalCcH{} and \RecRevalCcHPct\%. Across all \nCompared{} paired comparisons, \reval{} matched exhaustive reassessment while requiring \costSaving\% fewer post-comment test-case executions. Maintenance overhead increased wall-clock runtime for the microsecond-scale tests; the estimated break-even test cost was approximately \breakEven{}\,ms. These findings support revision-aware evidence maintenance within the constructed setting, while distinguishing revalidation correctness from witness adequacy. A pre-specified protocol provides a basis for evaluating generalisability to naturally occurring review histories.
\end{abstract}


% ================================================================== 1 INTRODUCTION
\section{Introduction}\label{sec:intro}

Modern code review is an established practice for improving software quality in industrial and open-source development \citep{bacchelli2013expectations,sadowski2018google,mcintosh2016empirical}. Its benefits extend beyond defect detection to include improvements in maintainability and evolvability \citep{beller2014problems,panichella2020relevant}. Research on review automation has progressed from pre-trained models for generating comments and refining code \citep{tufano2021towards,li2022codereviewer} to deployed systems that propose repairs, check development practices, and participate in pull-request workflows \citep{frommgen2024resolving,vijayvergiya2024autocommenter,cihan2025automated}. More recently, automated support has extended to determining whether previously reported concerns have been addressed \citep{github2026autoresolve}. This development introduces a distinct verification problem: what evidence justifies declaring a review concern resolved, and under what conditions does that judgement remain valid?

The distinction between administrative resolution and substantive repair is central to this problem. Review platforms record workflow events, including thread closure and the displacement of comments from their original code locations. These events describe the progression of a review, but do not, by themselves, establish that the underlying concern has been corrected. Nevertheless, resolution status is also used to evaluate automated review systems. Industrial reports characterise assistant impact through the proportion of automated comments that were resolved \citep{cihan2025automated}, while empirical studies operationalise resolution using platform thread state \citep{cynthia2026gohome}. Interpreting such measures as evidence of successful repair requires an assumption that administrative state corresponds to the status of the underlying concern. The validity of this assumption is therefore consequential for both automated decision-making and empirical evaluation.

A further difficulty arises from the temporal nature of software development. A repair that is correct at one revision may be invalidated by subsequent changes. These changes need not affect the originally commented lines: modifications to a helper function, configuration setting, or other dependency may reintroduce the concern while leaving the earlier resolution decision unchanged. Resolution should therefore be understood as a claim about a particular code revision. Maintaining that claim requires determining whether the evidence supporting it remains applicable as the code evolves.

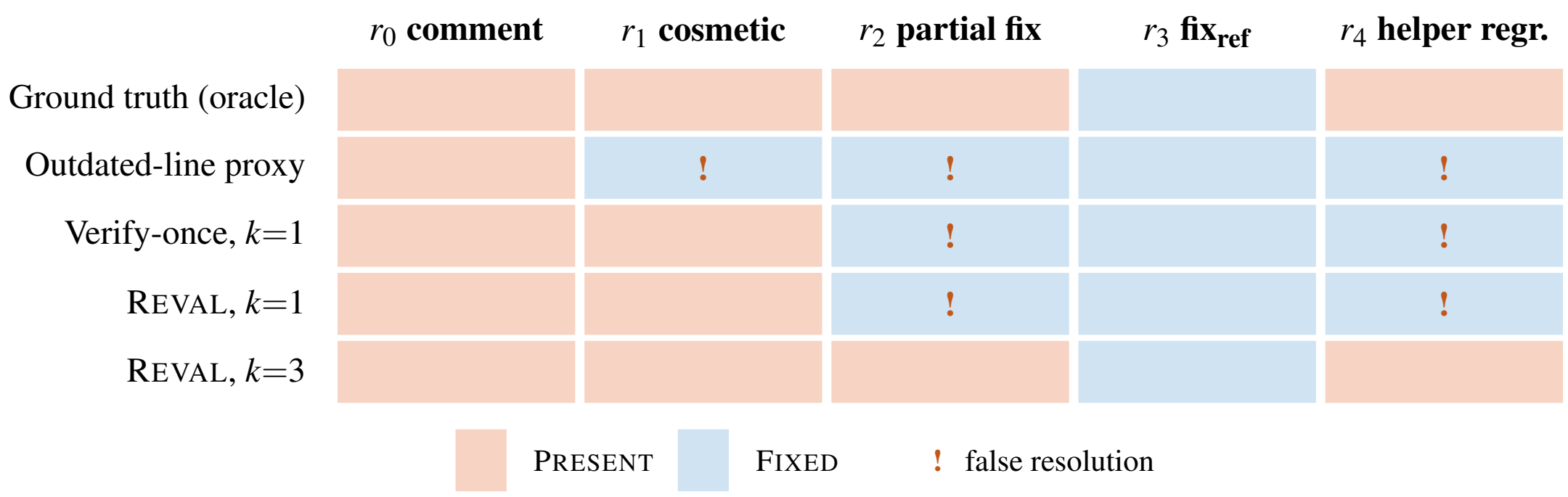
\begin{figure}[t]
\centering
\begin{tikzpicture}[x=2.25cm,y=0.62cm,
  cell/.style={draw=white,line width=1pt,minimum width=2.2cm,minimum height=0.6cm,font=\footnotesize,inner sep=0pt},
  lab/.style={font=\footnotesize,anchor=east}]
\foreach \i/\t in {0/{$r_0$ comment},1/{$r_1$ cosmetic},2/{$r_2$ partial fix},3/{$r_3$ fix\textsubscript{ref}},4/{$r_4$ helper regr.}}
  \node[font=\footnotesize\bfseries] at (\i,1) {\t};
\newcommand{\row}[7]{%
  \node[lab] at (-0.55,#1) {#2};
  \node[cell,fill=#3] at (0,#1) {}; \node[cell,fill=#4] at (1,#1) {};
  \node[cell,fill=#5] at (2,#1) {}; \node[cell,fill=#6] at (3,#1) {}; \node[cell,fill=#7] at (4,#1) {};}
\row{0}{Ground truth (oracle)}{presentc}{presentc}{presentc}{fixedc}{presentc}
\row{-1}{Outdated-line proxy}{presentc}{fixedc}{fixedc}{fixedc}{fixedc}
\row{-2}{Verify-once, $k{=}1$}{presentc}{presentc}{fixedc}{fixedc}{fixedc}
\row{-3}{\reval{}, $k{=}1$}{presentc}{presentc}{fixedc}{fixedc}{fixedc}
\row{-4}{\reval{}, $k{=}3$}{presentc}{presentc}{presentc}{fixedc}{presentc}
\foreach \y/\i/\m in {-1/1/!,-1/2/!,-1/4/!,-2/2/!,-2/4/!,-3/2/!,-3/4/!}
  \node[font=\footnotesize\bfseries,text=accent] at (\i,\y) {\m};
\node[cell,fill=presentc,minimum width=0.5cm] at (0.1,-5.3) {}; \node[font=\scriptsize,anchor=west] at (0.25,-5.3) {\Pres};
\node[cell,fill=fixedc,minimum width=0.5cm] at (1.0,-5.3) {}; \node[font=\scriptsize,anchor=west] at (1.15,-5.3) {\Fix};
\node[font=\scriptsize\bfseries,text=accent] at (1.95,-5.3) {!}; \node[font=\scriptsize,anchor=west] at (2.0,-5.3) {false resolution};
\end{tikzpicture}
\caption{The running example, built from the \texttt{quantity} family's variants. Each cell is the decision computed by running the method online. The $k{=}3$ witness set is the reviewer's example plus the cases \texttt{'0'} and \texttt{'7'}. The outdated-line proxy resolves at the cosmetic edit. Evidence from the reviewer's example alone accepts the partial fix. Only a maintained, stronger witness set tracks the true sequence.}
\label{fig:example}
\end{figure}

Consider a reviewer who observes at revision $r_0$ that \texttt{parse\_quantity('-3')} returns $-3$, although all non-positive quantities should raise \texttt{ValueError}. At $r_1$, reformatting makes the comment outdated without changing the behaviour. At $r_2$, a guard rejects negative values but continues to accept zero, producing a partial repair. At $r_3$, validation is moved to a helper with the correct boundary condition, fully addressing the concern. At $r_4$, a change to that helper causes zero to be accepted again. An outdated-line proxy registers a workflow transition at $r_1$, whereas a check restricted to the reviewer's example passes from $r_2$ onwards. Neither establishes that the complete concern is absent only at $r_3$ (\cref{fig:example}). The example distinguishes three sources of error: substituting workflow state for repair evidence, using evidence that captures only part of a concern, and retaining a resolution judgement after its supporting evidence has become invalid.

This paper advances the position that \emph{a resolution decision should be tied to a specific code revision and supported by inspectable evidence; when subsequent changes invalidate that evidence, the concern must be reassessed}. Under this position, review concerns are \emph{persistent verification obligations} whose status is maintained throughout a revision history. Two requirements follow. First, the evidence must adequately represent the concern being assessed. Second, its reuse at later revisions must be justified by the stability of the dependencies on which it relies. These requirements distinguish the strength of the evidence from the correctness of the procedure used to maintain it.

The position comprises three claims:
\begin{enumerate}[label=(\textbf{P\arabic*})]
  \item \emph{Measurement validity.} Administrative and syntactic proxies, including outdated-line status and changed-line heuristics, can diverge from the substantive repair of a review concern.
  \item \emph{Temporal validity.} A resolution judgement that is correct at one revision can become incorrect at a later revision. Methods that cease checking after the first successful assessment cannot detect subsequent recurrence.
  \item \emph{Evidence maintenance.} Under a sufficient dependency model, revalidating evidence when its dependencies change can preserve the decisions obtained by exhaustive reassessment while reducing executions. Agreement with exhaustive reassessment does not, however, establish that the evidence fully captures the concern.
\end{enumerate}

The computational principles underlying evidence maintenance are established. Safe regression-test selection identifies tests that require re-execution after a change \citep{rothermel1997safe,gligoric2015ekstazi}; build systems reuse results while their dependency fingerprints remain valid \citep{mokhov2018build}; and truth-maintenance systems revise beliefs when their justifications no longer hold \citep{doyle1979tms}. The contribution of this paper lies in applying and evaluating these principles at the level of persistent review concerns. A concern expresses an obligation that may be only partially represented by its executable witnesses. Evaluating its resolution consequently requires separating witness adequacy, dependency validity, and the policy governing reassessment.

We instantiate this perspective as \reval{}, a lightweight procedure for maintaining revision-specific resolution evidence. The procedure records the dependencies of an assessment, reuses its result while those dependencies remain unchanged, and triggers reassessment when relevant changes occur. Its evaluation follows individual concerns across revision histories, explicitly accounting for partial repairs and recurrence. The study addresses three research questions:
\begin{description}[leftmargin=1.2em,labelindent=0pt]
  \item[RQ1] In controlled revision histories, how frequently, and under which conditions, do an outdated-line proxy and a surface-pattern heuristic disagree with oracle-verified concern status?
  \item[RQ2] To what extent does maintaining revision-specific evidence reduce false resolution and improve recurrence detection relative to stopping after first success, and what is its cost relative to reassessment at every revision?
  \item[RQ3] How do witness strength and dependency granularity affect the remaining errors, including when witnesses are derived solely from review comments and evaluated against independently written checks?
\end{description}

The paper makes five contributions:
\begin{enumerate}[label=(\textbf{C\arabic*})]
  \item A \emph{formal model} of review concerns as revision-indexed verification obligations, including a sufficient condition for dependency-gated revalidation to match exhaustive reassessment (\cref{prop:sound}), an explicit account of the guarantee's limits (\cref{rem:levels}), and a decomposition of false resolution into three mechanisms (\cref{cor:decomp}).
  \item A \emph{taxonomy} characterising divergence between administrative closure and substantive repair (\cref{tab:taxonomy}).
  \item \reval{}, an \emph{evidence-maintenance procedure} with an explicit dependency model, conservative fallbacks, and regression tests for edge cases (\cref{sec:reval}).
  \item A \emph{controlled, executable evaluation} using oracle-sampled and blind comment-derived witnesses, with the latter assessed against independently written held-out checks. The evaluation includes shared-draw bootstrap uncertainty, separate witness-sampling uncertainty, and runtime measurements that distinguish execution costs from maintenance overhead (\cref{sec:study,sec:results}).
  \item A \emph{pre-specified protocol} for evaluating naturally occurring pull-request histories (\cref{sec:protocol}), supported by a replication package containing verification checks.
\end{enumerate}

The empirical scope is deliberately bounded. All quantitative findings are obtained from synthetic revision histories constructed from small, deterministic Python programs. The baselines isolate mechanisms through which resolution judgements can fail; they do not constitute measurements of GitHub, Copilot, or developer behaviour. The study therefore evaluates these mechanisms and their relative effects within the constructed setting, without estimating their prevalence in operational review workflows. The protocol for naturally occurring histories specifies the subsequent evaluation needed to assess external validity. Accordingly, the paper retains a position-paper framing: it establishes the conceptual and formal basis for revision-aware concern resolution and provides controlled evidence supporting further empirical investigation.

% ================================================================== 2 RELATED WORK
\section{Background and Related Work}\label{sec:related}

\subsection{Modern Code Review and Review Outcomes}

Empirical research has established that code review serves several complementary purposes. Although defect detection is a prominent motivation, field studies also identify maintainability improvements, knowledge transfer, and shared awareness as important outcomes \citep{bacchelli2013expectations,sadowski2018google}. Repository-mining studies have characterised the problems corrected during review \citep{beller2014problems}, examined associations between review practices and post-release defects \citep{mcintosh2016empirical}, and identified review-related changes that developers consider suitable for automation \citep{panichella2020relevant}. Related work has modelled comment usefulness using developer judgements and observable comment characteristics \citep{bosu2015useful,turzo2024useful}.

These studies establish the value and diversity of review outcomes, but evaluating whether a comment is useful differs from determining whether its underlying concern has been repaired. A useful comment may lead to an incomplete repair, and an initially successful repair may subsequently be reversed. A particularly relevant temporal perspective is provided by \citet{oliveira2025understandability}, who trace understandability-improvement comments through accepted changes and later reversions. Their analysis examines the persistence of review-induced changes. The present paper addresses the related but distinct question of whether a behavioural concern is absent at each subsequent revision, including cases in which the implementation changes without an explicit reversion of the original repair.

\subsection{Automation of Code Review and Comment Resolution}

Early approaches to review automation used pre-trained models to generate comments and refine code in response to review feedback \citep{tufano2021towards,li2022codereviewer}. Subsequent qualitative analysis identified recurring limitations underlying their apparent strengths \citep{tufano2024strengths}. Industrial deployments have extended this work to concrete review activities. At Google, authors addressed 7.5\% of all reviewer comments by applying an ML-suggested edit \citep{frommgen2024resolving}. Other systems support LLM-based best-practice checking \citep{vijayvergiya2024autocommenter} and automated participation in pull-request review; a deployment at Beko reported that 73.8\% of automated comments were resolved \citep{cihan2025automated}.

Agent-generated development activity has also enabled datasets such as AIDev \citep{li2025aidev}. Using AIDev-derived repositories, \citet{cynthia2026gohome} examine developer responses to agent-generated review comments and operationalise resolution through thread state. Automatic comment-resolution functionality extends automation to the resolution decision itself \citep{github2026autoresolve}.

These developments make the interpretation of resolution increasingly important. Applying a suggested edit, closing a thread, and establishing that a concern has been repaired are distinct outcomes. Reported adoption or resolution rates describe the corresponding workflow events, but their interpretation as evidence of substantive repair requires additional validation. Our study examines this distinction through controlled mechanism baselines. It makes no assumption that the cited systems implement those baselines or share their failure rates.

\subsection{Evaluation of Automated Review}

Evaluation methods for automated review address several different constructs. A mapping study documents heterogeneity in datasets and metrics \citep{wang2021benchmark}, while a recent roadmap advocates evaluation that accounts for context and practical value \citep{yang2026roadmap}. EvaCRC models attributes of review-comment quality \citep{yang2023evacrc}; CodeReviewQA evaluates comprehension of review intent \citep{lin2025codereviewqa}; and HalluJudge identifies generated comments that are insufficiently grounded in the code \citep{hallujudge2026}. These approaches assess properties of a comment or a model's interpretation of it.

Other evaluations incorporate executable evidence or iterative development. The c-CRAB benchmark derives tests from human reviews to assess agent-generated reviews \citep{ccrab2026}, while SWE-Review evaluates agentic reviewers through the resolve rate of generate--review--revise loops \citep{swereview2026}. Repository-level issue-resolution benchmarks provide a broader precedent for evaluating software changes against executable criteria \citep{jimenez2024swebench}. LLM-as-judge research offers a further methodological basis for assessments that rely on model-generated judgements \citep{zheng2023judging}.

The present paper complements these approaches by changing the unit and temporal scope of evaluation. Its unit is an individual review concern, and its outcome is the status of that concern at each revision. This formulation distinguishes initial repair from partial repair and subsequent recurrence. It also separates two questions that an aggregate success measure may combine: whether the available evidence is strong enough to represent the concern, and whether that evidence remains valid after later changes. \Cref{tab:positioning} summarises these distinctions for selected related studies.

\begin{table}[t]
\centering
\small
\caption{Positioning of selected related studies by evaluation focus. \checkmark{} denotes explicit treatment, $\circ$ denotes partial treatment, and -- denotes a dimension outside the stated evaluation focus.}
\label{tab:positioning}
\begin{tabularx}{\linewidth}{@{}l>{\raggedright\arraybackslash}Xccccc@{}}
\toprule
Work &
Unit of evaluation &
\makecell{Per\\concern} &
\makecell{Across\\revisions} &
\makecell{Partial\\repairs} &
\makecell{Recur-\\rence} &
\makecell{Executed\\evidence} \\
\midrule
EvaCRC \citep{yang2023evacrc}
& Comment quality attributes
& -- & -- & -- & -- & -- \\

CodeReviewQA \citep{lin2025codereviewqa}
& Comprehension of review intent
& \checkmark & -- & -- & -- & -- \\

HalluJudge \citep{hallujudge2026}
& Grounding of generated comments
& \checkmark & -- & -- & -- & -- \\

c-CRAB \citep{ccrab2026}
& Agent reviews assessed against human-derived tests
& \checkmark & -- & -- & -- & \checkmark \\

SWE-Review \citep{swereview2026}
& Resolution in review--revise loops
& -- & $\circ$ & -- & -- & \checkmark \\

\citet{oliveira2025understandability}
& Persistence of understandability changes
& \checkmark & \checkmark & -- & $\circ$ & -- \\

\citet{cynthia2026gohome}
& Developer responses and thread state
& \checkmark & -- & -- & -- & -- \\

\textbf{This paper}
& Concern status at each revision
& \checkmark & \checkmark & \checkmark & \checkmark & \checkmark \\
\bottomrule
\end{tabularx}
\end{table}

\subsection{Artefact Obsolescence and Evidence Validity}

Research on code--comment inconsistency examines how documentation that was initially accurate becomes inconsistent with evolving implementations \citep{wen2019inconsistencies}. Just-in-time inconsistency detection addresses this problem at the point of change \citep{panthaplackel2021jit}. This literature provides a useful conceptual parallel for review resolution: an assessment may be justified when made but become unsupported as the code changes.

The object maintained in this paper is the evidential basis of a resolution judgement. A review thread may retain its resolved status even after a change invalidates the checks that justified closure. Conversely, a substantial code change need not invalidate a judgement if the relevant evidence remains applicable. The resulting problem is therefore one of maintaining the relationship among a concern, its supporting evidence, and the revision to which that evidence applies.

\subsection{Dependency-Based Revalidation and Checkable Evidence}

Several established approaches provide the computational foundations for maintaining evidence under change. Safe regression-test selection determines which tests require re-execution following a program modification \citep{rothermel1997safe}. Dynamic dependency tracking supports selective execution using dependencies observed during earlier runs, including file-level tracking \citep{gligoric2015ekstazi}. Build systems similarly reuse derived results when their inputs and dependency fingerprints remain unchanged \citep{mokhov2018build}. Truth-maintenance systems provide a related account of how derived beliefs should be revised when their justifications cease to hold \citep{doyle1979tms}.

\reval{} adapts these principles to persistent review concerns. The concern remains the object of assessment, executable witnesses provide evidence about its status, and dependency information determines when that evidence requires revalidation. The formal guarantee concerns agreement with exhaustive execution of the same witnesses under the stated dependency assumptions. It does not establish that those witnesses constitute a complete specification of the concern. This distinction is essential because a sound revalidation policy can consistently preserve the result of an inadequate check.

Two additional research traditions inform the evaluation. Proof-carrying code motivates associating a decision with evidence that can be independently checked \citep{necula1997pcc}. Selective prediction motivates comparing error rates at matched decision coverage when methods may abstain from issuing a judgement \citep{geifman2017selective}. Together, these foundations support the paper's emphasis on inspectable evidence, explicit conditions for reuse, and evaluation that distinguishes decision reliability from the frequency with which decisions are made.
% ================================================================== 3 MODEL
\section{Concerns as Revision-Indexed Verification Obligations}\label{sec:model}

\subsection{Programs, revisions and concerns}

Let $\mathcal{H} = \langle r_0, \dots, r_n\rangle$ be the revision history of a change under review. Here $r_0$ is the commented revision and each $r_t$ is a complete program snapshot. A revision consists of modules. $\mathcal{U}(r)$ is its set of \emph{units}: the top-level functions and module-level bindings of every module, plus one \emph{namespace unit} $\nu_M$ per module $M$, whose text is the set of names bound at top level in $M$. We write $\mathrm{txt}_r(u)$ for the text of unit $u$ in $r$, with $\mathrm{txt}_r(u)=\bot$ if $u\notin\mathcal{U}(r)$. Namespace units make the \emph{addition} or \emph{removal} of a binding observable even when no existing unit's text changes.

\begin{definition}[Concern]
A \emph{concern} is $c = (\iota_c, A_c, r_0)$, with intent $\iota_c$ and anchor $A_c$ (the commented lines of $r_0$). The intent is made operational as an \emph{oracle} $\Omega_c = \{o_1,\dots,o_m\}$ of behavioural checks, $o_i(r) \in \{\top, \bot, \star\}$. Here $\star$ means the check is inapplicable because the subject of the concern cannot be located in $r$, and $o_1$ is the reviewer's own example.
\end{definition}

\begin{definition}[Concern status]
\begin{equation}
\sigma_c(r) =
\begin{cases}
\Und & \text{if } \exists i: o_i(r) = \star,\\
\Fix & \text{if } \forall i: o_i(r) = \top,\\
\Pres & \text{otherwise.}
\end{cases}
\end{equation}
Humans assign two further dispositions, which are never counted as repairs. A concern is \textsc{Invalid} if $\iota_c$ was wrong at $r_0$, and \textsc{Obsolete} if a later requirement change makes it inapplicable.
\end{definition}

A concern \emph{recurs} at $r_t$ if $\sigma_c(r_t) = \Pres$ and $\exists s<t: \sigma_c(r_s)=\Fix$. Status is a function of the revision, not a property of the thread.

\subsection{Resolution claims and their evidence}

\begin{definition}[Evidence-carrying claim]
A claim is $\kappa = (s, r_a, W, \rho, F, \phi)$. It consists of an asserted status $s$, an assessed revision $r_a$, a witness set $W$ of executed checks with outcomes $\rho$, a footprint $F \subseteq \mathcal{U}(r_a)$ of units the execution depended on, and a fingerprint $\phi(u) = \mathrm{txt}_{r_a}(u)$ for $u\in F$. The witness set need not be drawn from $\Omega_c$. In practice it is written from the review comment, and $\Omega_c$ is unknown.
\end{definition}

\begin{definition}[Evidence validity]
Claim $\kappa$ is \emph{valid at} $r$ if $s \ne \Und$ and $\forall u \in F: \mathrm{txt}_r(u) = \phi(u)$; otherwise it is \emph{stale at} $r$.
\end{definition}

\begin{proposition}[Soundness of dependency-gated revalidation]\label{prop:sound}
Assume that (i) each check in $W$ is deterministic, and (ii) $F$ is \emph{complete} for $W$: every revision that agrees with $r_a$ on the text of all units in $F$ yields the same outcomes for $W$. If $\kappa$ is valid at $r$, then re-executing $W$ on $r$ reproduces $\rho$, and hence $s$.
\end{proposition}
\begin{proof}
Validity gives agreement on $F$. Completeness then gives identical outcomes, and $s$ is a function of the outcomes.
\end{proof}

\begin{remark}[Three different levels of guarantee]\label{rem:levels}
\emph{(1) Mathematical.} \Cref{prop:sound} states \emph{equivalence to re-running the same witnesses}. It says nothing about whether passing $W$ implies $\sigma_c(r)=\Fix$, which depends on how well $W$ covers $\Omega_c$ (\cref{cor:decomp}b).
\emph{(2) Implemented.} Completeness is an assumption. Our implementation guarantees it only for the supported dependency model of \cref{sec:reval}: top-level functions, module-level bindings read by executed code, and module namespaces of executed modules, all for deterministic code. Dependencies outside this model, such as I/O, environment, time, randomness or reflection, require a conservative fallback.
\emph{(3) Empirical.} The agreement between \reval{} and exhaustive re-checking reported in \cref{sec:results} is an observation on this benchmark. It is consistent with completeness there but does not prove it.
\end{remark}

\begin{corollary}[Decomposition of false resolution]\label{cor:decomp}
If a method asserts \Fix{} at $r_t$ while $\sigma_c(r_t) = \Pres$, at least one of the following holds:
(a) \emph{no evidence}: the assertion is not backed by executing any check;
(b) \emph{witness insufficiency}: $W$ passes at $r_t$ but some check of $\Omega_c$ fails;
(c) \emph{stale evidence}: the claim was produced at $r_a<r_t$ and was not revalidated, either because the policy stopped checking or because $F$ was incomplete.
\end{corollary}
\begin{proof}
Suppose the assertion is backed by evidence. If that evidence was produced at $r_t$, or is valid at $r_t$ with $F$ complete (\cref{prop:sound}), then $W$ passes at $r_t$, which is case (b). Otherwise it is case (c).
\end{proof}

\subsection{Metrics}

Let $\mathcal{R}$ be the assessed (concern, revision) pairs with $t\ge 1$ and determinate truth. The primary metric is
\begin{equation}\label{eq:frr}
\FRR = \frac{\bigl|\{(c,r)\in\mathcal{R} : \sigma_c(r)=\Pres \wedge \hat\sigma_c(r)=\Fix\}\bigr|}{\bigl|\{(c,r)\in\mathcal{R} : \sigma_c(r)=\Pres\}\bigr|}.
\end{equation}
We also report fixed-label precision $P(\sigma=\Fix \mid \hat\sigma=\Fix)$, macro-$F_1$ over $\{\Fix,\Pres\}$, and \emph{recurrence recall} (recurring states predicted \Pres). We report the \emph{contract-change flag rate} (\Und{} states not asserted \Fix), \emph{decision coverage} (determinate states without abstention), and \emph{post-comment test-case executions} per concern, which exclude the initial assessment at $r_0$. Wall-clock runtime is reported separately (\cref{sec:runtime}).

\begin{table}[t]
\centering\small
\caption{Taxonomy of divergence between administrative closure and actual repair, with the mechanism of \cref{cor:decomp}. Rows above the rule are instantiated in the controlled study.}
\label{tab:taxonomy}
\begin{tabularx}{\linewidth}{@{}l>{\raggedright\arraybackslash}X>{\raggedright\arraybackslash}p{2.6cm}@{}}
\toprule
Divergence & What happens & Mechanism \\
\midrule
Cosmetic edit & Commented lines change, so the comment becomes outdated, but behaviour is unchanged & (a) no evidence\\
Partial fix & Reviewer's example passes; other in-scope cases still fail & (b) witness insufficiency\\
Relocated fix & Repair moves the check to a helper; line heuristics mis-score a correct fix & (a) for line proxies\\
Inline regression & A later edit of the same function drops the repair & (c) stale (policy)\\
Remote regression & A helper in another file is weakened; the commented function is untouched & (c) stale; needs footprint\\
Configuration regression & A module-level constant changes and re-opens the defect & (c) incomplete footprint\\
Contract change & The subject is renamed and the evidence cannot run; it may later be restored & should yield \Und\\
\midrule
Invalid concern & The original comment was wrong & human disposition\\
Obsolete concern & Requirements changed and the concern no longer applies & human disposition\\
\bottomrule
\end{tabularx}
\end{table}

% ================================================================== 4 REVAL
\section{\reval{}: Revision-Aware Evidence Maintenance}\label{sec:reval}

\reval{} keeps one \emph{concern record} per atomic concern. A record holds the intent and anchor, the witness set $W$, the latest claim $\kappa$ and the history of past claims. A reviewer can therefore see \emph{which revision} supports the current status and \emph{why} it changed. \Cref{alg:reval} gives the online procedure.

\begin{algorithm}[t]
\small
\caption{\reval{}: online assessment of concern $c$ over revisions}\label{alg:reval}
\KwIn{history $r_0,\dots,r_n$ (revealed one at a time); witness set $W$; granularity $g\in\{\textsf{file},\textsf{func},\textsf{func+globals}\}$}
\KwOut{status $\hat\sigma_c(r_t)$ and supporting claim for each $t$}
$\kappa \leftarrow \textsc{Assess}(r_0, W, g)$\;
\For{$t \leftarrow 1$ \KwTo $n$}{
  \eIf{$\kappa.s = \Und$ \textbf{or} $\exists u \in \kappa.F:\ \mathrm{txt}_{r_t}(u) \ne \kappa.\phi(u)$}{
     $\kappa \leftarrow \textsc{Assess}(r_t, W, g)$ \tcp*{no valid evidence: reassess}
  }{
     keep $\kappa$ \tcp*{evidence valid: decision carries over}
  }
  emit $(\kappa.s, \kappa.r_a)$\;
}
\SetKwFunction{FA}{Assess}\SetKwProg{Fn}{Function}{:}{}
\Fn{\FA{$r, W, g$}}{
  run each check of $W$ on a fresh instance of $r$ under a call tracer; collect outcomes $\rho$, executed functions $E$ (and their modules $M_E$), and names $N$ read by executed code\;
  $s \leftarrow \Und$ if any check is inapplicable; else \Fix{} if all pass; else \Pres\;
  \uIf{$g=\textsf{func+globals}$}{$F \leftarrow E \,\cup\, \{\text{module-level bindings named in } N\} \,\cup\, \{\nu_M : M\in M_E\}$\;}
  \uElseIf{$g=\textsf{func}$}{$F\leftarrow E$\;}
  \lElse{$F \leftarrow$ the modules $M_E$, compared by whole-file text}
  \Return $(s, r, \rho, F, \phi_r|_F)$\;
}
\end{algorithm}

\paragraph{Supported dependency model and fallbacks.}
The footprint is \emph{dynamic}, collected while the witnesses run. Repairs that move logic into helpers therefore extend $F$ automatically. Three edge cases are handled explicitly. First, while a concern is \Und{} there is no valid evidence, so \emph{every} new revision is reassessed (line~3). This covers restoration of a missing target by \emph{adding} a definition, which leaves all existing units unchanged. Second, namespace units detect new bindings, for example a module-level name that shadows a builtin read by the target, and removed bindings. Third, a dependency outside the model (I/O, environment, time, randomness, reflection, dynamic import) cannot be traced. In that case the record must fall back to file-level or every-revision reassessment. In this benchmark every witness is deterministic and in-model, so no fallback was triggered. The implementation ships with regression tests for target removal, restoration by addition, builtin shadowing, helper and constant changes, and unrelated edits.

\paragraph{Assessor-agnostic.}
Here $W$ consists of executable checks. The same record and staleness test can also gate re-invocation of an LLM assessor, with the footprint taken from the code the assessor was shown. That variant belongs to the protocol in \cref{sec:protocol}.

% ================================================================== 5 STUDY
\section{Controlled Study Design}\label{sec:study}

\subsection{Benchmark construction}

We constructed \nFamilies{} \emph{concern families} (\cref{tab:families}, \cref{app:families}). Each is a two-file Python program, with \texttt{core.py} holding the commented function and constants and \texttt{util.py} holding helpers. Each comes with a review comment and a hidden oracle of 6--7 behavioural checks, including positive cases that a correct fix must keep passing. The families cover input validation (3), security (path traversal, SQL injection, output escaping), exception handling (2), resource management, boundary conditions, null handling and shared mutable state. Each family provides a defective commented version, a partial fix, a complete inline fix, a complete fix delegating to a helper and a weakened helper. Three families also have a configuration constant whose regression re-opens the defect. By design, partial fixes and weakened helpers make the reviewer's example pass while other in-scope checks fail.

Histories come from \nScen{} scenario scripts (\cref{tab:scenarios}), each instantiated with three seeds that insert zero to two unrelated edits at random positions. This yields \nConcerns{} concerns and \nAssess{} post-comment assessments, with \revPerConcern{} revisions per concern on average. Ground truth is obtained by \emph{executing the full oracle} at every revision. It contains \nPresent{} \Pres{}, \nFixed{} \Fix{} and \nUndet{} \Und{} states, including \nRecur{} recurring states. The \Pres{} states comprise cosmetic edits (\nCatCos), partial fixes (\nCatPart), untouched (\nCatUnt), remote helper regressions (\nCatHelp), inline regressions (\nCatInl) and configuration regressions (\nCatConst).

\begin{table}[t]
\centering\small
\caption{Scenario scripts (unrelated edits are inserted at random). ``fix\textsubscript{ref}'' is a complete fix delegating to a helper; ``restore'' in S13 re-adds the original function name by a new definition.}
\label{tab:scenarios}
\begin{tabular}{@{}lll@{}}
\toprule
ID & Developer actions after the comment & Probes \\
\midrule
S01 & fix & baseline repair\\
S02 & cosmetic $\to$ fix & premature closure\\
S03 & partial & witness insufficiency\\
S04 & partial $\to$ fix & delayed true repair\\
S05 & fix $\to$ inline regression & local recurrence\\
S06 & fix\textsubscript{ref} $\to$ helper regression & remote recurrence\\
S07 & fix $\to$ constant regression (3 families) & configuration recurrence\\
S08 & cosmetic $\to$ cosmetic & never repaired\\
S09 & fix $\to$ behaviour-preserving refactor & control (no status change)\\
S10 & fix\textsubscript{ref} $\to$ helper regression $\to$ restore helper & recurrence and re-repair\\
S11 & cosmetic $\to$ partial & compound divergence\\
S12 & fix $\to$ rename target & contract change (\Und)\\
S13 & fix $\to$ rename target $\to$ restore by addition & re-anchoring after \Und\\
\bottomrule
\end{tabular}
\end{table}

\subsection{Methods compared}\label{sec:methods}

All methods run online and see $r_t$ only after deciding on $r_{t-1}$. The first two are \emph{controlled mechanism baselines}. Each isolates one failure mechanism and is not a model of any product's behaviour.
\begin{itemize}
  \item \textbf{Outdated-line proxy} (\textsc{Out}). It predicts \Fix{} from the first revision at which the commented lines no longer appear verbatim, and the decision is \emph{sticky} by construction. It represents using ``the comment became outdated'' (or a thread resolved on that basis) as a repair label. It is \emph{not} the decision procedure of GitHub's \texttt{isOutdated} handling or of Copilot's auto-resolution, whose assessment method is not public \citep{github2026autoresolve}.
  \item \textbf{Surface-pattern heuristic} (\textsc{Pat}). A stateless AST heuristic: \Fix{} if the commented function contains more defensive constructs (conditionals, raises, \texttt{try}/\texttt{with}, sanitising or clamping calls, helper delegation) than at $r_0$. It represents judgements that reward fixes that merely \emph{look} defensive. It is not an LLM judge.
  \item \textbf{Evidence policies.} \textsc{Once} assesses at $r_0$, then only when the commented function's text changes while the status is not yet \Fix{}. Once \Fix{} is asserted, it \emph{stops checking by design}, so its zero recurrence recall follows from the policy. \textsc{Always} reassesses at every revision. \textsc{File}, \textsc{Func} and \reval{} (\textsc{Func+Globals}) reassess when their footprint is stale (\cref{alg:reval}). \textsc{Always} and \textsc{Once} run without the tracer.
\end{itemize}

\subsection{Witness conditions}\label{sec:witnesses}

\paragraph{Experiment A: oracle-sampled witnesses.}
With $k=1$, the witness is the reviewer's example ($o_1$). With $k=3$, it is $o_1$ plus two checks sampled from $\Omega_c$ without replacement, independently for each concern. Each concern is evaluated under \nDraws{} independent draws, and every oracle has at least six checks, so no budget $k\le 5$ exhausts it; the code asserts this. This condition is a controlled ablation. It gives the method well-chosen checks from the same source as the ground truth, so it shows what maintenance achieves \emph{given} strong witnesses.

\paragraph{Experiment B: blind comment-derived witnesses and held-out ground truth.}
To evaluate the approach on concerns where no oracle is available, two further authors wrote checks under blinding. Each author was a separate AI agent instance with no conversation history. Each saw only the review comment and the commented revision of \texttt{core.py} for every family, was instructed not to open any other file, and never saw the oracle, the variants or any future revision. The \emph{witness author} wrote exactly three checks per family, as a review assistant would, with the first reproducing the reviewer's example. The \emph{held-out author} independently wrote 6--7 acceptance checks per family that define ``fully fixed without breaking legitimate behaviour''. Both files were frozen (SHA-256 recorded) before any evaluation and were not edited afterwards. In Experiment B, ground truth $\sigma^{H}$ is the held-out suite. The witnesses are scored against it and, for comparison, against the original oracle. Blinding was enforced by instruction and by giving each agent only the input file, not by sandboxing. Both authors are instances of the same model family as the one that assisted in preparing this study. They are therefore independent of the oracle's content but not independent human judges (\cref{sec:threats}).

\subsection{Statistical analysis and runtime}

Assessments within a family share code and checks, so all confidence intervals are 95\% percentile intervals from a \emph{family-clustered bootstrap} \citep{cameron2008bootstrap} with \nBoot{} resamples of the \nFamilies{} families. \emph{One} matrix of resampled family indices is shared by all methods and all paired differences. Methods with identical predictions therefore receive identical intervals, and paired differences are directly comparable. Two sources of uncertainty are kept separate. Family selection is captured by the bootstrap of the draw-averaged metric. Witness sampling is captured by the range of the point estimate across the \nDraws{} draws. With twelve clusters, intervals indicate stability across concern types, not population inference.

Runtime is measured as wall-clock time per concern, median of \timingRepeats{} repeats on Python \pyVersion{}, with the $k{=}3$ witnesses of draw 0. It is split into execution (compiling, loading and running checks, tracer included where active), fingerprinting (parsing units), staleness comparison and remaining policy logic.

% ================================================================== 6 RESULTS
\section{Results}\label{sec:results}

\begin{table}[t]
\centering\small
\setlength{\tabcolsep}{4.2pt}
\caption{Experiment A (oracle-sampled witnesses) over \nAssess{} post-comment assessments of \nConcerns{} concerns. Prec\textsubscript{F}: fixed-label precision; Rec\textsubscript{rec}: recurrence recall; Flag\textsubscript{U}: contract-change states not asserted \Fix; Exec.: post-comment test-case executions per concern. Brackets: 95\% family-clustered bootstrap CIs (shared draws). $k{=}3$ rows average \nDraws{} witness draws. Decision coverage on determinate states is 1.00 for all methods.}
\label{tab:main}
\begin{tabular}{@{}lcccccc@{}}
\toprule
Method & $\FRR\downarrow$ & Prec\textsubscript{F}$\uparrow$ & Macro-$F_1\uparrow$ & Rec\textsubscript{rec}$\uparrow$ & Flag\textsubscript{U}$\uparrow$ & Exec.$\downarrow$\\
\midrule
\multicolumn{7}{l}{\emph{Mechanism baselines (no execution)}}\\
\quad Outdated-line proxy & 0.73 {\scriptsize[0.66, 0.78]} & 0.53 & 0.54 & 0.05 {\scriptsize[0.00, 0.16]} & 0.12 & 0.00 \\
\quad Surface-pattern judge & 0.37 {\scriptsize[0.33, 0.41]} & 0.69 & 0.76 & 0.36 {\scriptsize[0.29, 0.48]} & 1.00 & 0.00 \\
\addlinespace
\multicolumn{7}{l}{\emph{Evidence, reviewer's example only ($k{=}1$)}}\\
\quad Verify-once & 0.48 {\scriptsize[0.46, 0.49]} & 0.65 & 0.74 & 0.00 {\scriptsize[0.00, 0.00]} & 0.00 & 1.11 \\
\quad Always recheck & 0.37 {\scriptsize[0.33, 0.40]} & 0.71 & 0.80 & 0.45 {\scriptsize[0.33, 0.59]} & 1.00 & 2.91 \\
\quad File-level footprint & 0.37 {\scriptsize[0.33, 0.40]} & 0.71 & 0.80 & 0.45 {\scriptsize[0.33, 0.59]} & 1.00 & 2.53 \\
\quad Function footprint & 0.37 {\scriptsize[0.33, 0.40]} & 0.70 & 0.80 & 0.42 {\scriptsize[0.31, 0.56]} & 1.00 & 1.89 \\
\quad \textbf{REVAL} (functions + globals) & 0.37 {\scriptsize[0.33, 0.40]} & 0.71 & 0.80 & 0.45 {\scriptsize[0.33, 0.59]} & 1.00 & 1.91 \\
\addlinespace
\multicolumn{7}{l}{\emph{Evidence, oracle-sampled witness set ($k{=}3$; mean of \nDraws{} draws)}}\\
\quad Verify-once & 0.30 {\scriptsize[0.27, 0.34]} & 0.75 & 0.84 & 0.00 {\scriptsize[0.00, 0.00]} & 0.00 & 3.51 \\
\quad Always recheck & 0.11 {\scriptsize[0.06, 0.16]} & 0.89 & 0.94 & 0.80 {\scriptsize[0.73, 0.88]} & 1.00 & 8.73 \\
\quad File-level footprint & 0.11 {\scriptsize[0.06, 0.16]} & 0.89 & 0.94 & 0.80 {\scriptsize[0.73, 0.88]} & 1.00 & 7.60 \\
\quad Function footprint & 0.12 {\scriptsize[0.07, 0.18]} & 0.88 & 0.93 & 0.73 {\scriptsize[0.63, 0.85]} & 1.00 & 5.67 \\
\quad \textbf{REVAL} (functions + globals) & 0.11 {\scriptsize[0.06, 0.16]} & 0.89 & 0.94 & 0.80 {\scriptsize[0.73, 0.88]} & 1.00 & 5.73 \\
\bottomrule
\end{tabular}

\end{table}

\subsection{RQ1: Mechanism baselines diverge from repair in structured ways}

\begin{figure}[t]
\centering
\includegraphics[width=0.78\linewidth]{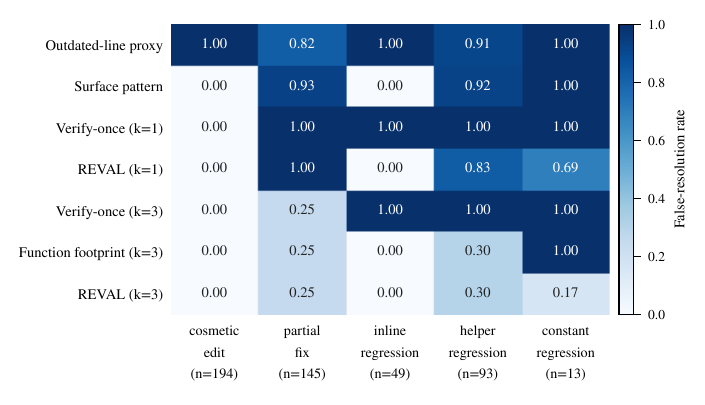}
\caption{False-resolution rate by divergence type (columns) and method (rows), Experiment A. Each method fails on a characteristic subset, as \cref{cor:decomp} predicts.}
\label{fig:frr-cat}
\end{figure}

In the synthetic histories, the sticky outdated-line proxy falsely marked \outClosedPresent{} of \nPresent{} still-present concern states as resolved ($\FRR=\FrrOut$, CI \FrrOutLo--\FrrOutHi; \cref{tab:main}). It disagreed with oracle status at least once in \outDivergeConcerns\% of histories, and its fixed-label precision was \PrecOut. The errors follow the mechanisms (\cref{fig:frr-cat}). Every cosmetic edit, inline regression and configuration regression produced a false resolution. Errors also ran the other way: on \outOpenFixed{} fixed states the commented line survived unchanged because the fix added lines around it. These magnitudes follow from the proxy's construction and the scenario mix, not from measured platform behaviour. What generalises is the \emph{pattern}: a line-change signal cannot separate cosmetic from behavioural edits, nor detect regressions after it fires.

The surface-pattern heuristic reached $\FRR=\FrrPat$ and failed in complementary places. It was immune to cosmetic edits and inline regressions, which remove defensive constructs, but accepted \CatPartPat{} of partial fixes and \CatHelpPat{} of remote regressions, because these look like fixes or leave the commented function unchanged.

\begin{figure}[t]
\centering
\includegraphics[width=\linewidth]{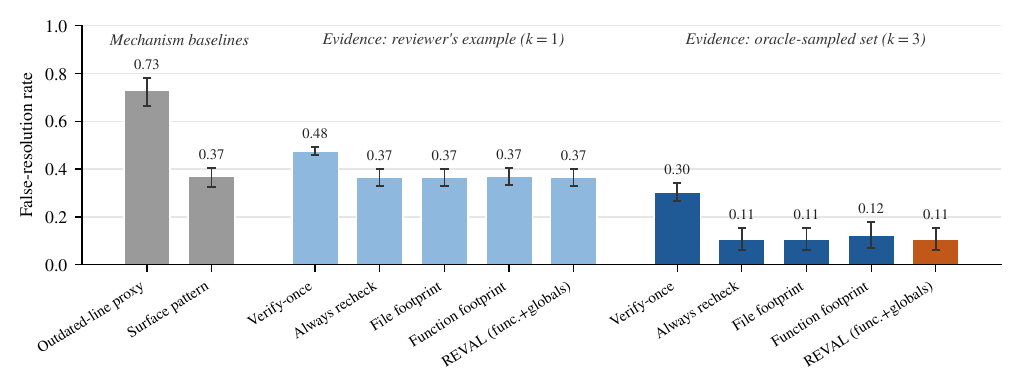}
\caption{False-resolution rate in Experiment A with 95\% family-clustered bootstrap CIs (shared draws). Grey: mechanism baselines; light blue: reviewer's example only; dark blue: oracle-sampled $k{=}3$ (mean of \nDraws{} draws); orange: \reval{}.}
\label{fig:frr}
\end{figure}

\subsection{RQ2: Maintaining evidence detects recurrence at fewer executions}

\textsc{Once} reduced false resolution relative to the proxy but, by design, had zero recurrence recall. With the same $k{=}3$ witnesses, \reval{} reduced the false-resolution rate from \FrrOnceK{} to \FrrRevalK{} (paired difference \DFrrOnce, CI \DFrrOnceLo{} to \DFrrOnceHi). It raised recurrence recall to \RecRevalK{} (CI \RecRevalKLo--\RecRevalKHi). Across witness draws, its FRR ranged from \FrrRevalKDrawLo{} to \FrrRevalKDrawHi.

\begin{figure}[t]
\centering
\includegraphics[width=0.56\linewidth]{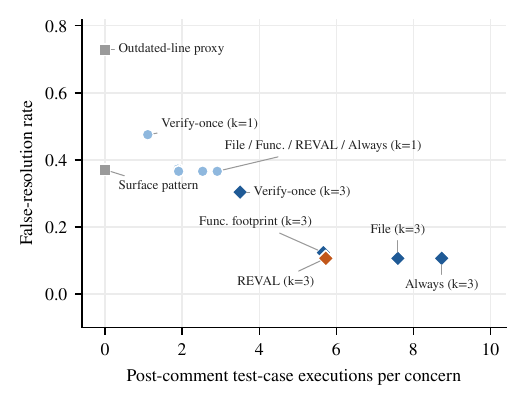}
\caption{Accuracy versus post-comment test-case executions (Experiment A). Squares: baselines; circles: $k{=}1$; diamonds: $k{=}3$. \reval{} (orange) matches \textsc{Always} with fewer executions. Wall-clock runtime is a different matter (\cref{sec:runtime}).}
\label{fig:cost}
\end{figure}

\reval{}'s decisions were identical to \textsc{Always} in all \nCompared{} paired assessments (\agreeRevalAlways\%), as \cref{prop:sound} predicts when the footprint is complete. It needed \CostRevalK{} instead of \CostAlwaysK{} post-comment test-case executions per concern, \costSaving\% fewer (\cref{fig:cost}). Unrelated edits rarely triggered reassessment (\reassUnrelReval\% of unrelated-only revisions, all of them while a concern was \Und{}), against \reassUnrelFile\% for file-level footprints. Contract changes were flagged \Und{} by every re-checking policy. In S13, re-adding the target by a new definition was always detected and reassessed. The sticky policies carried their earlier \Fix{} over (Flag\textsubscript{U} column).

\subsection{RQ3: Dependency granularity and witness strength govern residual error}

\paragraph{Granularity.}
Function-only footprints missed every configuration regression (\cref{fig:frr-cat}: \CatConstFuncK{} vs.\ \CatConstRevalK). This lowered recurrence recall by \DRecFunc{} (CI \DRecFuncLo--\DRecFuncHi) and agreement with \textsc{Always} to \agreeFuncAlways\%. Adding globals read by executed code, together with namespace units, restored full agreement at \DCostFunc{} extra executions per concern. File-level footprints also agreed fully but reassessed on most unrelated edits.

\begin{figure}[t]
\centering
\includegraphics[width=0.48\linewidth]{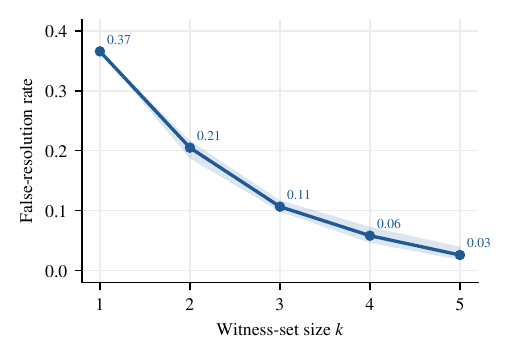}\hfill
\includegraphics[width=0.48\linewidth]{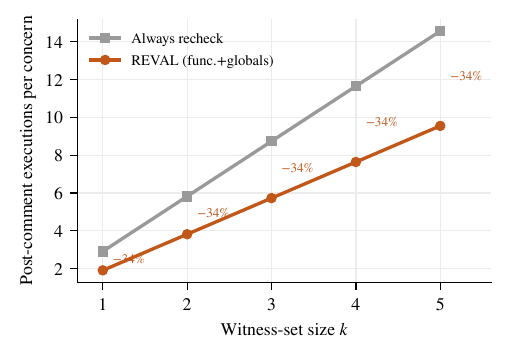}
\caption{Witness-set size ablation under \reval{}, oracle-sampled witnesses (same sampler as \cref{tab:main}; band: range over \nDraws{} draws). Left: false-resolution rate. Right: post-comment executions for \reval{} and \textsc{Always}. Decisions of the two policies were identical at every $k$.}
\label{fig:sweep}
\end{figure}

\paragraph{Witness strength.}
With only the reviewer's example, every evidence policy accepted all partial fixes. This is a witness-insufficiency floor ($\FRR\approx\FrrRevalS$) that no revalidation removes, and its size partly reflects our construction of partial fixes. \Cref{fig:sweep} varies $k$ using \emph{the same sampler and draws} as \cref{tab:main}, so its $k{=}3$ point (\sweepFrrc) equals the table entry. Under \reval{}, FRR falls from \sweepFrra{} ($k{=}1$) through \sweepFrrb{} and \sweepFrrc{} to \sweepFrrd{} and \sweepFrre{} ($k{=}4,5$), and recurrence recall rises from \sweepReca{} to \sweepRece. No budget exhausts an oracle, so $k{=}5$ still leaves at least one hidden check. At every $k$, \reval{} matched \textsc{Always}, with \sweepSaveMin--\sweepSaveMax\% fewer post-comment executions. This stability is an \emph{observation}, not a guarantee. Extra witnesses can enlarge the dynamic footprint and change which revisions require reassessment. Here they did not, because all witnesses of a family exercise the same functions.

\subsection{Witnesses written without oracle access (Experiment B)}\label{sec:indep}

\Cref{tab:indep} reports Experiment B, separately from Experiment A. The two blind ground truths largely agree: the held-out suite and the original oracle gave the same status on \truthAgree\% of assessments. The \nTruthDisagree{} disagreements are concentrated in three families where the comment underdetermines intent. Is mutating a caller-supplied list acceptable in \texttt{add\_tag}? Must non-transient errors skip retries in \texttt{with\_retry}? Does a ``dotted domain'' exclude a trailing dot in \texttt{normalize\_email}? Specification ambiguity is thus a real source of disagreement even before any assessor is involved.

Against the held-out truth, \reval{} with all three comment-derived witnesses reached $\FRR=\FrrRevalCcH$ (CI \FrrRevalCcHLo--\FrrRevalCcHHi). Its recurrence recall was \RecRevalCcH{}, and it never rejected a held-out-fixed state (missed-repair rate \MissRevalCcH). That compares with \FrrRevalCaH{} using only the first comment-derived witness (difference \DFrrSingleC, CI \DFrrSingleCLo{} to \DFrrSingleCHi) and \FrrOnceCcH{} for \textsc{Once} with the same witnesses (difference \DFrrOnceC, CI \DFrrOnceCLo{} to \DFrrOnceCHi). Again \reval{} matched \textsc{Always} on every decision (\agreeRevalAlwaysCmt\%). Scored against the original oracle instead, the same method reached $\FRR=\FrrRevalCcO$. The remaining false resolutions came from three families. In \texttt{mean} and \texttt{display\_name}, the witness author did not test a sibling input (an empty tuple, an empty profile), which is witness insufficiency. In \texttt{normalize\_email}, the two blind authors read the comment differently, which is specification disagreement rather than an assessment error. In this setting, then, comment-derived witnesses performed comparably to oracle-sampled ones of the same size (\FrrRevalKH{} against the held-out truth). The confidence intervals are wide.

\begin{table}[t]
\centering\small
\setlength{\tabcolsep}{4.2pt}
\caption{Experiment B: witnesses written blind from the review comment and initial code, scored against an independently, blind-written held-out suite ($\sigma^{H}$: \nPresentH{} \Pres, \nFixedH{} \Fix{}, \nRecurH{} recurring states). Columns and CIs as in \cref{tab:main}. Oracle-sampled rows are re-scored against $\sigma^{H}$ for reference.}
\label{tab:indep}
\begin{tabular}{@{}lcccccc@{}}
\toprule
Method & $\FRR\downarrow$ & Prec\textsubscript{F}$\uparrow$ & Macro-$F_1\uparrow$ & Rec\textsubscript{rec}$\uparrow$ & Flag\textsubscript{U}$\uparrow$ & Exec.$\downarrow$\\
\midrule
\multicolumn{7}{l}{\emph{Mechanism baselines}}\\
\quad Outdated-line proxy & 0.73 {\scriptsize[0.65, 0.80]} & 0.52 & 0.53 & 0.06 {\scriptsize[0.00, 0.21]} & 0.12 & 0.00 \\
\quad Surface-pattern judge & 0.35 {\scriptsize[0.27, 0.41]} & 0.70 & 0.79 & 0.36 {\scriptsize[0.30, 0.47]} & 1.00 & 0.00 \\
\addlinespace
\multicolumn{7}{l}{\emph{Blind comment-derived witnesses, first check ($k{=}1$)}}\\
\quad Verify-once & 0.46 {\scriptsize[0.36, 0.56]} & 0.65 & 0.74 & 0.00 {\scriptsize[0.00, 0.00]} & 0.00 & 1.12 \\
\quad Always recheck & 0.36 {\scriptsize[0.22, 0.49]} & 0.71 & 0.81 & 0.52 {\scriptsize[0.37, 0.70]} & 1.00 & 2.91 \\
\quad \textbf{REVAL} (functions + globals) & 0.36 {\scriptsize[0.22, 0.49]} & 0.71 & 0.81 & 0.52 {\scriptsize[0.37, 0.70]} & 1.00 & 1.91 \\
\addlinespace
\multicolumn{7}{l}{\emph{Blind comment-derived witnesses, all three ($k{=}3$)}}\\
\quad Verify-once & 0.32 {\scriptsize[0.23, 0.43]} & 0.73 & 0.83 & 0.00 {\scriptsize[0.00, 0.00]} & 0.00 & 3.51 \\
\quad Always recheck & 0.13 {\scriptsize[0.00, 0.26]} & 0.87 & 0.93 & 0.88 {\scriptsize[0.72, 1.00]} & 1.00 & 8.73 \\
\quad \textbf{REVAL} (functions + globals) & 0.13 {\scriptsize[0.00, 0.26]} & 0.87 & 0.93 & 0.88 {\scriptsize[0.72, 1.00]} & 1.00 & 5.73 \\
\addlinespace
\multicolumn{7}{l}{\emph{Reference: oracle-sampled witnesses ($k{=}3$)}}\\
\quad Verify-once & 0.35 {\scriptsize[0.26, 0.44]} & 0.71 & 0.80 & 0.00 {\scriptsize[0.00, 0.00]} & 0.00 & 3.51 \\
\quad \textbf{REVAL} (functions + globals) & 0.17 {\scriptsize[0.06, 0.29]} & 0.83 & 0.88 & 0.82 {\scriptsize[0.73, 0.91]} & 1.00 & 5.73 \\
\bottomrule
\end{tabular}

\end{table}

\subsection{Runtime and overhead}\label{sec:runtime}

Execution counts are not runtime. \Cref{tab:timing} separates the components. In this benchmark a single check runs in about \perExecUntraced\,$\mu$s untraced and \perExecTraced\,$\mu$s under Python's call tracer, a factor of \traceFactor. Fingerprinting dominates \reval{}'s cost. As a result, \reval{} was \timeRatio$\times$ \emph{slower} in wall-clock time than untraced exhaustive re-checking, despite \costSaving\% fewer executions. Its overhead per concern (\revalOverhead\,ms) is repaid only when one test execution costs more than about \breakEven\,ms. Real test suites usually exceed that by orders of magnitude, and LLM assessments exceed it far more, but we did not measure such settings. The efficiency claim of this paper is therefore limited to \emph{fewer post-comment test-case executions}. Total computational savings depend on the cost of an assessment relative to tracing and fingerprinting.

\begin{table}[t]
\centering\small
\caption{Wall-clock runtime per concern (median of \timingRepeats{} repeats, $k{=}3$, draw 0), including the initial assessment at $r_0$. Execution includes the tracer where active.}
\label{tab:timing}
\begin{tabular}{@{}lrrrrrr@{}}
\toprule
Policy & Exec./concern & Execution & Fingerprint & Staleness & Other & Total\\
& (incl.\ $r_0$) & ms & ms & ms & ms & ms\\
\midrule
Always recheck (untraced) & 11.73 & 0.40 & 0.00 & 0.00 & 0.10 & 0.50 \\
Verify-once (untraced) & 6.49 & 0.25 & 0.00 & 0.00 & 0.25 & 0.50 \\
File-level footprint & 10.60 & 0.85 & 0.46 & 0.01 & 0.15 & 1.46 \\
\textbf{REVAL} & 8.73 & 0.75 & 0.73 & 0.01 & 0.16 & 1.64 \\
\bottomrule
\end{tabular}

\end{table}

\subsection{Summary: support within the constructed setting}

Within the synthetic histories, the mechanism baselines diverged from repair in structured ways (P1). Policies that stop after first success detected no recurrence, while maintained evidence did (P2). Dependency-gated revalidation reproduced exhaustive re-checking with fewer executions, provided the footprint covered globals and namespaces, and its residual error was governed by witness strength and, in Experiment B, by specification ambiguity (P3). These are mechanism-level results. Their prevalence and their payoff on real repositories remain to be measured.

% ================================================================== 7 DISCUSSION
\section{Discussion}\label{sec:discussion}

\subsection{Implications for tool builders}

\paragraph{Resolve with a receipt.}
An automated resolution should record the revision assessed, the checks or reasoning used, and what that judgement depended on. This makes ``resolved'' auditable, in the spirit of proof-carrying code \citep{necula1997pcc}.

\paragraph{Resolution is not terminal.}
A resolved concern should re-open when its evidence goes stale, as a build system invalidates a cached result \citep{mokhov2018build}. The footprint must include module-level state and namespaces, not only functions.

\paragraph{Abstain when the subject moves, and reassess until re-anchored.}
Renames should yield an explicit ``needs re-anchoring'' state, which is reassessed at every revision until valid evidence exists again.

\paragraph{Spend effort on witnesses; a proposed workflow.}
Residual error was dominated by witness insufficiency and specification ambiguity. We therefore \emph{propose}, without having evaluated it, a workflow in which an assistant drafts boundary and sibling cases from the comment and the reviewer \emph{confirms or edits} them, so that ambiguities such as those in \cref{sec:indep} are resolved by the person who raised the concern. Experiment B evaluated only the first half, assistant-drafted witnesses without reviewer confirmation. The confirmation step, its cost to reviewers and its effect on accuracy are untested.

\paragraph{Mind the overhead.}
Where assessments are cheap, re-checking everything may be faster than maintaining footprints (\cref{sec:runtime}). Evidence maintenance pays off where assessments are expensive, such as integration tests or LLM-based judgements, and where the receipt itself has value.

\subsection{Implications for researchers}

Thread state (\texttt{isResolved}, \texttt{isOutdated}) is an \emph{observed workflow variable}, not a correctness label. Reported ``resolution rates'' of AI-generated comments \citep{cihan2025automated,cynthia2026gohome} measure tool and developer behaviour, not the repair rate of the underlying concerns. Benchmarks for review agents \citep{ccrab2026,swereview2026} would gain from scoring concerns longitudinally and at matched decision coverage \citep{geifman2017selective}. They should also report agreement between independently specified ground truths, since our two blind specifications disagreed on \nTruthDisagree{} of \nAssess{} states.

\subsection{Protocol for naturally occurring histories}\label{sec:protocol}

We defer a natural-history study and pre-specify its protocol (\cref{tab:protocol}).

\paragraph{Corpus.}
Start from AIDev \citep{li2025aidev}, whose review tables cover pull requests in repositories with more than 100 stars. Reconstruct histories from the GitHub review-comment and review-thread APIs. Record PR authorship and comment authorship separately. Begin with Python and behaviourally checkable concerns. Audit 50 histories for reconstruction reliability, then target about 300 atomic concerns across at least 15 repositories.

\paragraph{Ground truth.}
Two human annotators independently label each concern's validity and its per-revision status. Agreement is reported before adjudication \citep{cohen1960kappa}. A 30--50-case runnable subset receives independently written tests \citep{ccrab2026}.

\paragraph{Assessors and controls.}
Compare thread state and outdated status as observed (not simulated), a per-revision LLM judge, the same judge with full history, and the judge gated by a \reval{}-style record. Use at least two model families with matched information and budgets, an ablation without the record, hidden future revisions and repository hold-outs. Measure wall-clock and monetary cost alongside assessment counts.

\paragraph{Falsification.}
The position is weakened in three cases. First, if observed thread state agrees closely with verified repair (P1). Second, if recurrence after correct resolution is too rare for post-closure checking to matter (P2). Third, if evidence records do not reduce FRR over an equally informed per-revision judge at matched coverage and cost (P3).

\begin{table}[t]
\centering\small
\caption{Pre-specified protocol for the natural-history study.}
\label{tab:protocol}
\begin{tabularx}{\linewidth}{@{}l>{\raggedright\arraybackslash}X>{\raggedright\arraybackslash}X@{}}
\toprule
Experiment & Design & Establishes \\
\midrule
E1 Proxy audit & Observed thread closure and outdated status vs.\ adjudicated per-revision status & Prevalence of each divergence (P1)\\
E2 Assessor comparison & Per-revision vs.\ history-aware vs.\ evidence-maintaining LLM judges, matched budgets & Whether evidence helps beyond context (P3)\\
E3 Stress sequences & Cosmetic, partial, complete, recurrence and control edits built from real concerns & Tracking of the underlying concern (P2)\\
E4 Executable subset & Independent tests on 30--50 runnable cases & Label agreement with behaviour\\
E5 Witness workflow & Assistant-drafted witnesses with and without reviewer confirmation & Value and cost of confirmation\\
\bottomrule
\end{tabularx}
\end{table}

\subsection{Threats to validity}\label{sec:threats}

\paragraph{Construct validity.}
Each oracle encodes one reading of a comment. Experiment B shows that a second, blind reading agrees on \truthAgree\% of states, and that the disagreements are interpretable ambiguities. The outdated-line proxy is a simplification of GitHub's diff-position logic, and it is sticky by design. It should be read as a mechanism baseline, not as an estimate of platform or developer error.

\paragraph{Internal validity.}
We designed the code variants, so magnitudes partly reflect design choices. In particular, partial fixes pass the reviewer's example by construction. Ground truth is always computed by execution. All methods are online and deterministic. The blind authors in Experiment B are AI agent instances, blinded by instruction rather than sandboxing, and from the same model family that assisted with the study. Their independence is therefore from the oracle's content, not from shared model biases. A human-authored replication is part of the protocol.

\paragraph{External validity.}
Programs are small, single-language and deterministic. Many real concerns, such as readability or design, have no executable witness. Real dependencies include I/O, configuration files and non-determinism outside our dependency model. Runtime results depend on the cost of an assessment relative to tracing, and we measured only microsecond-scale checks.

\paragraph{Conclusion validity.}
With twelve clusters, bootstrap intervals are approximate. Family-selection and witness-sampling uncertainty are reported separately. Key contrasts are large relative to both, but small differences should not be over-interpreted.

% ================================================================== 8 CONCLUSION
\section{Conclusion}\label{sec:conclusion}

As AI assistants begin to decide when review concerns are resolved, the field needs a definition of resolution that survives code evolution. We proposed one: a resolution is a revision-specific, evidence-carrying claim that must be revalidated when what it depends on changes.

\paragraph{Key findings (synthetic histories).}
(1) A sticky outdated-line proxy falsely marked \FrrOutPct\% of still-present concern states as resolved, failing on cosmetic edits, partial fixes and all regressions. (2) A policy that stops after first success detected no recurrence. (3) \reval{} reduced false resolution to \FrrRevalK{} with oracle-sampled witnesses and to \FrrRevalCcH{} with blind comment-derived witnesses scored against an independent held-out suite. It detected most recurrences and matched exhaustive re-checking on every decision, with \costSaving\% fewer post-comment test-case executions. (4) Residual error came from witness insufficiency and specification ambiguity. Function-only footprints additionally missed configuration regressions.

\paragraph{Limitations.}
The evidence is synthetic and covers behavioural concerns in small deterministic programs. Blind authors were AI agents, not humans. No LLM assessor was evaluated. With microsecond-scale checks, the approach was slower in wall-clock time than re-checking everything.

\paragraph{Implications.}
Tools should attach receipts to automated resolutions, re-open concerns when receipts go stale, abstain when the subject moves, and invest in witnesses confirmed by reviewers. Researchers should not treat thread state as a repair label, and should report agreement between independent specifications.

\paragraph{Future directions.}
Future work should (i) run the natural-history protocol, with human annotators; (ii) gate LLM assessors with evidence records and measure monetary cost; (iii) evaluate reviewer-confirmed witness drafting; and (iv) extend the dependency model to configuration files and I/O, with conservative fallbacks, inside continuous integration.

\section*{Data and Code Availability}
The replication package (\texttt{experiment/}) contains the following:
\begin{itemize}
  \item the concern families and oracles;
  \item the frozen blind-authored witness and held-out files, with their input file and SHA-256 hashes;
  \item every generated history, with the source code of each revision, ground truth and sampled witnesses (\texttt{results/histories.jsonl});
  \item raw per-assessment predictions (\texttt{results/rows.csv});
  \item pinned dependency versions and a single command (\texttt{reproduce.sh}) that regenerates all tables, figures and numbers;
  \item regression tests (\texttt{tests.py});
  \item \texttt{verify.py}, which checks the reported counts (\nConcerns{} histories, \nAssess{} assessments, class counts), \reval{}--\textsc{Always} agreement, sweep/table consistency and file hashes against \texttt{expected.json}.
\end{itemize}

\section*{Declarations}
\textbf{Funding:} none. \textbf{Competing interests:} none. \textbf{Use of AI tools:} AI assistance was used for drafting, code development and reference checking. Separate AI agent instances authored the blind witness and held-out checks of Experiment B, as described in \cref{sec:witnesses}. The author reviewed and takes responsibility for all content.

\bibliographystyle{plainnat}
\bibliography{refs}

@misc{github2026autoresolve,
  author       = {{GitHub}},
  title        = {Copilot code review: An improved review experience},
  howpublished = {GitHub Changelog},
  year         = {2026},
  month        = sep,
  day          = {18},
  url          = {https://github.blog/changelog/2026-09-18-copilot-code-review-an-improved-review-experience},
  note         = {Accessed 2026-10-20}
}

@article{tufano2024strengths,
  author  = {Tufano, Rosalia and Dabi\'{c}, Ozren and Mastropaolo, Antonio and Ciniselli, Matteo and Bavota, Gabriele},
  title   = {Code Review Automation: Strengths and Weaknesses of the State of the Art},
  journal = {IEEE Transactions on Software Engineering},
  volume  = {50},
  number  = {2},
  pages   = {338--353},
  year    = {2024},
  doi     = {10.1109/TSE.2023.3348172}
}

@article{oliveira2025understandability,
  author  = {Oliveira, Delano and Santos, Reydne and de Oliveira, Benedito and Monperrus, Martin and Castor, Fernando and Madeiral, Fernanda},
  title   = {Understanding Code Understandability Improvements in Code Reviews},
  journal = {IEEE Transactions on Software Engineering},
  volume  = {51},
  number  = {1},
  pages   = {14--37},
  year    = {2025},
  doi     = {10.1109/TSE.2024.3453783}
}

@article{turzo2024useful,
  author  = {Turzo, Asif Kamal and Bosu, Amiangshu},
  title   = {What makes a code review useful to {OpenDev} developers? {An} empirical investigation},
  journal = {Empirical Software Engineering},
  volume  = {29},
  number  = {1},
  pages   = {6},
  year    = {2024},
  doi     = {10.1007/s10664-023-10411-x}
}

@article{panichella2020relevant,
  author  = {Panichella, Sebastiano and Zaugg, Nik},
  title   = {An empirical investigation of relevant changes and automation needs in modern code review},
  journal = {Empirical Software Engineering},
  volume  = {25},
  number  = {6},
  pages   = {4833--4872},
  year    = {2020},
  doi     = {10.1007/s10664-020-09870-3}
}

@article{yang2026roadmap,
  author  = {Yang, Zezhou and Gao, Cuiyun and Guo, Zhaoqiang and Li, Zhenhao and Liu, Kui and Xia, Xin and Zhou, Yuming},
  title   = {A Roadmap for Modern Code Review: Challenges and Opportunities},
  journal = {ACM Transactions on Software Engineering and Methodology},
  volume  = {35},
  number  = {9},
  pages   = {1--42},
  year    = {2026},
  doi     = {10.1145/3800963},
  note    = {Preprint: arXiv:2405.18216}
}

@article{wang2021benchmark,
  author  = {Wang, Dong and Ueda, Yuki and Kula, Raula Gaikovina and Ishio, Takashi and Matsumoto, Kenichi},
  title   = {Can we benchmark Code Review studies? {A} systematic mapping study of methodology, dataset, and metric},
  journal = {Journal of Systems and Software},
  volume  = {180},
  pages   = {111009},
  year    = {2021},
  doi     = {10.1016/j.jss.2021.111009}
}

@inproceedings{hallujudge2026,
  author    = {Tantithamthavorn, Kla and Lin, Hong Yi and Thongtanunam, Patanamon and Charoenwet, Wachiraphan and Jeong, Minwoo and Wu, Ming},
  title     = {{HalluJudge}: A Reference-Free Hallucination Detection for Context Misalignment in Code Review Automation},
  booktitle = {Proceedings of the 34th ACM International Conference on the Foundations of Software Engineering (FSE Companion '26), Industry Track},
  pages     = {632--643},
  year      = {2026},
  publisher = {ACM},
  doi       = {10.1145/3803437.3805236},
  note      = {arXiv:2601.19072}
}

@inproceedings{lin2025codereviewqa,
  author    = {Lin, Hong Yi and Liu, Chunhua and Gao, Haoyu and Thongtanunam, Patanamon and Treude, Christoph},
  title     = {{CodeReviewQA}: The Code Review Comprehension Assessment for Large Language Models},
  booktitle = {Findings of the Association for Computational Linguistics: {ACL} 2025},
  pages     = {9138--9166},
  year      = {2025},
  address   = {Vienna, Austria},
  publisher = {Association for Computational Linguistics},
  doi       = {10.18653/v1/2025.findings-acl.476}
}

@inproceedings{yang2023evacrc,
  author    = {Yang, Lanxin and Xu, Jinwei and Zhang, Yifan and Zhang, He and Bacchelli, Alberto},
  title     = {{EvaCRC}: Evaluating Code Review Comments},
  booktitle = {Proceedings of the 31st ACM Joint European Software Engineering Conference and Symposium on the Foundations of Software Engineering (ESEC/FSE)},
  pages     = {275--287},
  year      = {2023},
  doi       = {10.1145/3611643.3616245}
}

@inproceedings{li2022codereviewer,
  author    = {Li, Zhiyu and Lu, Shuai and Guo, Daya and Duan, Nan and Jannu, Shailesh and Jenks, Grant and Majumder, Deep and Green, Jared and Svyatkovskiy, Alexey and Fu, Shengyu and Sundaresan, Neel},
  title     = {Automating Code Review Activities by Large-Scale Pre-training},
  booktitle = {Proceedings of the 30th ACM Joint European Software Engineering Conference and Symposium on the Foundations of Software Engineering (ESEC/FSE)},
  pages     = {1035--1047},
  year      = {2022},
  doi       = {10.1145/3540250.3549081}
}

@misc{ccrab2026,
  author       = {Zhang, Yuntong and Pan, Zhiyuan and Yusuf, Imam Nur Bani and Ruan, Haifeng and Shariffdeen, Ridwan and Roychoudhury, Abhik},
  title        = {Code Review Agent Benchmark},
  year         = {2026},
  eprint       = {2603.23448},
  archivePrefix= {arXiv},
  primaryClass = {cs.SE},
  howpublished = {arXiv preprint arXiv:2603.23448},
  url          = {https://arxiv.org/abs/2603.23448}
}

@misc{swereview2026,
  author       = {Wang, Ruoyu and Chen, Jierun and Wang, Shaowei and Tao, Chaofan and Yang, Sidi and Jiang, Yuxin and Yap, Kim-Hui and Shang, Lifeng and Li, Xiaohui and Bai, Haoli},
  title        = {{SWE-Review}: Closing the Loop on Issue Resolution with Agentic Code Review},
  year         = {2026},
  eprint       = {2607.06065},
  archivePrefix= {arXiv},
  primaryClass = {cs.SE},
  howpublished = {arXiv preprint arXiv:2607.06065},
  url          = {https://arxiv.org/abs/2607.06065}
}

@misc{li2025aidev,
  author       = {Li, Hao and Zhang, Haoxiang and Hassan, Ahmed E.},
  title        = {The Rise of {AI} Teammates in Software Engineering ({SE}) 3.0: How Autonomous Coding Agents Are Reshaping Software Engineering},
  year         = {2025},
  eprint       = {2507.15003},
  archivePrefix= {arXiv},
  primaryClass = {cs.SE},
  howpublished = {arXiv preprint arXiv:2507.15003. Dataset: \url{https://huggingface.co/datasets/hao-li/AIDev}},
  url          = {https://arxiv.org/abs/2507.15003},
  note         = {Accessed 2026-10-20}
}

@misc{cynthia2026gohome,
  author       = {Cynthia, Shamse Tasnim and Widyasari, Ratnadira and Roy, Banani and Zhang, Ting and Lo, David},
  title        = {``Go Home {Copilot}, You're Drunk'': Understanding Developer Responses to Agent-Generated Code Review Comments},
  year         = {2026},
  eprint       = {2607.21997},
  archivePrefix= {arXiv},
  primaryClass = {cs.SE},
  howpublished = {arXiv preprint arXiv:2607.21997},
  url          = {https://arxiv.org/abs/2607.21997}
}

@inproceedings{bacchelli2013expectations,
  author    = {Bacchelli, Alberto and Bird, Christian},
  title     = {Expectations, Outcomes, and Challenges of Modern Code Review},
  booktitle = {Proceedings of the 35th International Conference on Software Engineering (ICSE)},
  pages     = {712--721},
  year      = {2013},
  publisher = {IEEE},
  doi       = {10.1109/ICSE.2013.6606617}
}

@inproceedings{sadowski2018google,
  author    = {Sadowski, Caitlin and S\"{o}derberg, Emma and Church, Luke and Sipko, Michal and Bacchelli, Alberto},
  title     = {Modern Code Review: A Case Study at {Google}},
  booktitle = {Proceedings of the 40th International Conference on Software Engineering: Software Engineering in Practice (ICSE-SEIP)},
  pages     = {181--190},
  year      = {2018},
  publisher = {ACM},
  doi       = {10.1145/3183519.3183525}
}

@inproceedings{beller2014problems,
  author    = {Beller, Moritz and Bacchelli, Alberto and Zaidman, Andy and Juergens, Elmar},
  title     = {Modern Code Reviews in Open-Source Projects: Which Problems Do They Fix?},
  booktitle = {Proceedings of the 11th Working Conference on Mining Software Repositories (MSR)},
  pages     = {202--211},
  year      = {2014},
  publisher = {ACM},
  doi       = {10.1145/2597073.2597082}
}

@article{mcintosh2016empirical,
  author  = {McIntosh, Shane and Kamei, Yasutaka and Adams, Bram and Hassan, Ahmed E.},
  title   = {An empirical study of the impact of modern code review practices on software quality},
  journal = {Empirical Software Engineering},
  volume  = {21},
  number  = {5},
  pages   = {2146--2189},
  year    = {2016},
  doi     = {10.1007/s10664-015-9381-9}
}

@inproceedings{tufano2021towards,
  author    = {Tufano, Rosalia and Pascarella, Luca and Tufano, Michele and Poshyvanyk, Denys and Bavota, Gabriele},
  title     = {Towards Automating Code Review Activities},
  booktitle = {Proceedings of the 43rd IEEE/ACM International Conference on Software Engineering (ICSE)},
  pages     = {163--174},
  year      = {2021},
  doi       = {10.1109/ICSE43902.2021.00027}
}

@inproceedings{frommgen2024resolving,
  author    = {Fr\"{o}mmgen, Alexander and Austin, Jacob and Choy, Peter and Ghelani, Nimesh and Kharatyan, Lera and Surita, Gabriela and Khrapko, Elena and Lamblin, Pascal and Manzagol, Pierre-Antoine and Revaj, Marcus and Tabachnyk, Maxim and Tarlow, Daniel and Villela, Kevin and Zheng, Daniel and Chandra, Satish and Maniatis, Petros},
  title     = {Resolving Code Review Comments with Machine Learning},
  booktitle = {Proceedings of the 46th International Conference on Software Engineering: Software Engineering in Practice (ICSE-SEIP)},
  pages     = {204--215},
  year      = {2024},
  publisher = {ACM},
  doi       = {10.1145/3639477.3639746}
}

@inproceedings{vijayvergiya2024autocommenter,
  author    = {Vijayvergiya, Manushree and Salawa, Ma{\l}gorzata and Budiseli\'{c}, Ivan and Zheng, Dan and Lamblin, Pascal and Ivankovi\'{c}, Marko and Carin, Juanjo and Lewko, Mateusz and Andonov, Jovan and Petrovi\'{c}, Goran and Tarlow, Daniel and Maniatis, Petros and Just, Ren\'{e}},
  title     = {{AI}-Assisted Assessment of Coding Practices in Modern Code Review},
  booktitle = {Proceedings of the 1st ACM International Conference on AI-Powered Software (AIware)},
  pages     = {85--93},
  year      = {2024},
  publisher = {ACM},
  doi       = {10.1145/3664646.3665664}
}

@inproceedings{gligoric2015ekstazi,
  author    = {Gligoric, Milos and Eloussi, Lamyaa and Marinov, Darko},
  title     = {Practical Regression Test Selection with Dynamic File Dependencies},
  booktitle = {Proceedings of the 2015 International Symposium on Software Testing and Analysis (ISSTA)},
  pages     = {211--222},
  year      = {2015},
  publisher = {ACM},
  doi       = {10.1145/2771783.2771784}
}

@article{rothermel1997safe,
  author  = {Rothermel, Gregg and Harrold, Mary Jean},
  title   = {A Safe, Efficient Regression Test Selection Technique},
  journal = {ACM Transactions on Software Engineering and Methodology},
  volume  = {6},
  number  = {2},
  pages   = {173--210},
  year    = {1997},
  doi     = {10.1145/248233.248262}
}

@inproceedings{necula1997pcc,
  author    = {Necula, George C.},
  title     = {Proof-Carrying Code},
  booktitle = {Proceedings of the 24th ACM SIGPLAN-SIGACT Symposium on Principles of Programming Languages (POPL)},
  pages     = {106--119},
  year      = {1997},
  publisher = {ACM},
  doi       = {10.1145/263699.263712}
}

@article{doyle1979tms,
  author  = {Doyle, Jon},
  title   = {A Truth Maintenance System},
  journal = {Artificial Intelligence},
  volume  = {12},
  number  = {3},
  pages   = {231--272},
  year    = {1979},
  doi     = {10.1016/0004-3702(79)90008-0}
}

@article{mokhov2018build,
  author  = {Mokhov, Andrey and Mitchell, Neil and {Peyton Jones}, Simon},
  title   = {Build Systems \`{a} la Carte},
  journal = {Proceedings of the ACM on Programming Languages},
  volume  = {2},
  number  = {ICFP},
  pages   = {1--29},
  year    = {2018},
  doi     = {10.1145/3236774}
}

@inproceedings{zheng2023judging,
  author    = {Zheng, Lianmin and Chiang, Wei-Lin and Sheng, Ying and Zhuang, Siyuan and Wu, Zhanghao and Zhuang, Yonghao and Lin, Zi and Li, Zhuohan and Li, Dacheng and Xing, Eric P. and Zhang, Hao and Gonzalez, Joseph E. and Stoica, Ion},
  title     = {Judging {LLM}-as-a-Judge with {MT-Bench} and {Chatbot Arena}},
  booktitle = {Advances in Neural Information Processing Systems 36 (NeurIPS 2023), Datasets and Benchmarks Track},
  pages     = {46595--46623},
  year      = {2023}
}

@inproceedings{jimenez2024swebench,
  author    = {Jimenez, Carlos E. and Yang, John and Wettig, Alexander and Yao, Shunyu and Pei, Kexin and Press, Ofir and Narasimhan, Karthik},
  title     = {{SWE-bench}: Can Language Models Resolve Real-World {GitHub} Issues?},
  booktitle = {The Twelfth International Conference on Learning Representations (ICLR)},
  year      = {2024},
  url       = {https://openreview.net/forum?id=VTF8yNQM66}
}

@inproceedings{geifman2017selective,
  author    = {Geifman, Yonatan and El-Yaniv, Ran},
  title     = {Selective Classification for Deep Neural Networks},
  booktitle = {Advances in Neural Information Processing Systems 30 (NIPS 2017)},
  year      = {2017},
  url       = {https://proceedings.neurips.cc/paper/2017/hash/4a8423d5e91fda00bb7e46540e2b0cf1-Abstract.html}
}

@article{cameron2008bootstrap,
  author  = {Cameron, A. Colin and Gelbach, Jonah B. and Miller, Douglas L.},
  title   = {Bootstrap-Based Improvements for Inference with Clustered Errors},
  journal = {The Review of Economics and Statistics},
  volume  = {90},
  number  = {3},
  pages   = {414--427},
  year    = {2008},
  doi     = {10.1162/rest.90.3.414}
}

@article{cohen1960kappa,
  author  = {Cohen, Jacob},
  title   = {A Coefficient of Agreement for Nominal Scales},
  journal = {Educational and Psychological Measurement},
  volume  = {20},
  number  = {1},
  pages   = {37--46},
  year    = {1960},
  doi     = {10.1177/001316446002000104}
}

@inproceedings{bosu2015useful,
  author    = {Bosu, Amiangshu and Greiler, Michaela and Bird, Christian},
  title     = {Characteristics of Useful Code Reviews: An Empirical Study at {Microsoft}},
  booktitle = {Proceedings of the 12th IEEE/ACM Working Conference on Mining Software Repositories (MSR)},
  pages     = {146--156},
  year      = {2015},
  publisher = {IEEE},
  doi       = {10.1109/MSR.2015.21}
}

@inproceedings{cihan2025automated,
  author    = {Cihan, Umut and Haratian, Vahid and \.{I}\c{c}\"{o}z, Arda and G\"{u}l, Mert Kaan and Devran, \"{O}mercan and Bayendur, Emircan Furkan and U\c{c}ar, Baykal Mehmet and T\"{u}z\"{u}n, Eray},
  title     = {Automated Code Review In Practice},
  booktitle = {Proceedings of the 47th IEEE/ACM International Conference on Software Engineering: Software Engineering in Practice (ICSE-SEIP)},
  year      = {2025},
  note      = {arXiv:2412.18531},
  url       = {https://arxiv.org/abs/2412.18531}
}

@inproceedings{wen2019inconsistencies,
  author    = {Wen, Fengcai and Nagy, Csaba and Bavota, Gabriele and Lanza, Michele},
  title     = {A Large-Scale Empirical Study on Code-Comment Inconsistencies},
  booktitle = {Proceedings of the 27th IEEE/ACM International Conference on Program Comprehension (ICPC)},
  pages     = {53--64},
  year      = {2019},
  doi       = {10.1109/ICPC.2019.00019}
}

@inproceedings{panthaplackel2021jit,
  author    = {Panthaplackel, Sheena and Li, Junyi Jessy and Gligoric, Milos and Mooney, Raymond J.},
  title     = {Deep Just-In-Time Inconsistency Detection Between Comments and Source Code},
  booktitle = {Proceedings of the AAAI Conference on Artificial Intelligence},
  volume    = {35},
  pages     = {427--435},
  year      = {2021},
  doi       = {10.1609/aaai.v35i1.16119}
}

\appendix
\section{Concern Families}\label{app:families}

\begin{table}[h]
\centering\small
\caption{The twelve concern families. ``Oracle'' gives the number of hidden checks; $\dagger$ marks families with a configuration constant.}
\label{tab:families}
\begin{tabularx}{\linewidth}{@{}l l >{\raggedright\arraybackslash}X c@{}}
\toprule
Family & Category & Review comment & Oracle\\
\midrule
quantity$^\dagger$ & input validation & \texttt{parse\_quantity('-3')} returns $-3$; non-positive must raise & 7\\
mean & exception handling & \texttt{mean([])} raises \texttt{ZeroDivisionError}; return \texttt{None} & 7\\
paginate & boundary condition & \texttt{paginate(items,1,3)} skips the first page; page $<1$ invalid & 7\\
safe\_join & security & \texttt{'../etc/passwd'} escapes the base directory & 7\\
user\_query & security & name interpolated into SQL; use a bound parameter & 6\\
greeting & security & \texttt{'<script>'} emitted raw; escape user text & 6\\
first\_line & resource management & handle leaks when \texttt{readline} raises \texttt{OSError} & 6\\
with\_retry$^\dagger$ & exception handling & all errors swallowed, returns \texttt{None}; surface the error & 6\\
display\_name & null handling & crashes when \texttt{profile} is \texttt{None}; fall back to name & 6\\
set\_volume$^\dagger$ & input validation & \texttt{set\_volume(150)} returns 150; clamp to $[0,\mathrm{MAX}]$ & 7\\
add\_tag & state management & mutable default list leaks tags across calls & 6\\
normalize\_email & input validation & \texttt{'bob@'} accepted; need local part and dotted domain & 7\\
\bottomrule
\end{tabularx}
\end{table}

\section{Blind Authoring Protocol (Experiment B)}\label{app:blind}

Each author received one JSON file listing, per family, the family identifier, target function name, review comment and the commented revision of \texttt{core.py}. The witness author was asked to act as a review assistant and write exactly three checks per family, the first reproducing the reviewer's example, without over-specifying incidental details such as error messages. The held-out author was asked to write 6--8 acceptance checks per family covering the full scope of the concern, including positive cases. Both were told not to open any other file and could execute their checks only against the defective initial code. Both reported their resolutions of ambiguity. For example, both accepted any placeholder style for SQL parameters and supplied several candidate keys for the unspecified ``account name''. The held-out author read ``dotted domain'' strictly and did not require non-transient errors to skip retries. The instructions, the outputs and their hashes are part of the replication package.

\end{document}